\documentclass[11pt]{article}
\pdfoutput=1
\usepackage{amsmath,sint,cite,epsfig,graphics,macros_static}
\usepackage{mathrsfs}
\usepackage{amssymb}
\usepackage{multirow}
\begin{document}

\newcommand{\Vub}{V_{\rm ub}}
\newcommand{\Obound}[2]{{\cal O}_{#1}(#2)}
\newcommand{\Oboundp}[2]{{\cal O}'_{#1}(#2)}
\renewcommand{\fone}{F_{1}}
\renewcommand{\kone}{K_{1}}
\renewcommand{\fv}{F_{\rm V_0}}
\newcommand{\ja}{J^1_{\rm A_1}}
\newcommand{\Hone}{H^{(1)}}
\newcommand{\Hzero}{H^{(0)}}
\newcommand{\Gzero}{G^{(0)}}
\newcommand{\Gone}{G^{(1)}}
\newcommand{\pastor}{{\texttt{pastor}}}

\begin{titlepage}

\begin{flushright}
\small{
DESY 12-187 \\
SFB/CPP-12-81
}
\end{flushright}

\begin{center}
  {\Large\bf A one-loop study of matching conditions for static-light flavor currents
  }
\end{center}
\vskip 1 cm
\vbox{
\centerline{
\epsfxsize=2.5 true cm
\epsfbox{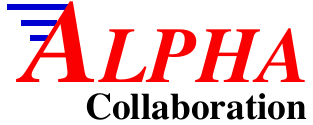}
} 
}
\vskip 0.5 cm
\begin{center}
{ Dirk Hesse$^{\scriptscriptstyle a,b}$ and Rainer Sommer$^{\scriptscriptstyle a}$
}
\vskip 3.5ex
$^{\scriptstyle a}$
        NIC, DESY, Platanenallee 6, 15738~Zeuthen,  Germany
\vskip 2.0ex
$^{\scriptstyle b}$
Universit\`{a} degli studi di Parma,\\
 Viale G.P. Usberti n.\ 7/A (Parco Area delle Scienze), 43124~Parma, Italy
\vskip 2.0ex

\vskip 0.5cm
{\bf Abstract}\\[.5cm]
\end{center}

  Heavy Quark Effective Theory (HQET) computations of
  semi-leptonic decays, e.g.\ $B\to\pi l \nu$, require
  the knowledge of the parameters in the effective
  theory for all components of the heavy-light flavor currents. So far
  non-perturbative matching conditions have been employed
  only for the time component of the axial current.   
  Here we perform a check of matching conditions for 
  the time component of the vector current and the spatial component
  of the axial vector current up to one-loop order of perturbation theory
  and to lowest order of the $1/m$-expansion. 
  We find that the proposed observables have 
  small higher order terms in the $1/m$-series and are thus
  excellent candidates for a non-perturbative matching procedure.
\vskip 0.7ex

\vskip 2.0ex
\noindent{\it Key words:}
Heavy Quark Effective Theory; Lattice QCD; Non-perturbative renormalization; Matching

\noindent{\it PACS:} 
12.38.Bx; 
12.38.Gc; 
12.39.Hg; 
13.20.He  
\vskip 2.0ex

\vfill
\eject

\end{titlepage}

\newcommand{\npar}{19 }
\newcommand{\mren}{\mbar}
\renewcommand{\zahqet}{Z_{\rm A_0}^{\rm HQET}}
\renewcommand{\zvhqet}{Z_{\rm V_0}^{\rm HQET}}
\renewcommand{\zakhqet}{Z_{\rm A_k}^{\rm HQET}}
\renewcommand{\zvkhqet}{Z_{\rm V_k}^{\rm HQET}}

\section{Introduction}

B meson decays are an excellent source of information
for constraining physics beyond the Standard Model. 
Precision based on a  solid theory and advanced experiments
is becoming increasingly important as we know that 
effects due to fields which are not present in the Standard Model 
are small. Next to leptonic decays, exclusive semileptonic
decays are easiest to treat in theory. Take for example the decay
$B \to \pi l\nu$ which is relevant for a determination
of $\Vub$. Theory only needs to predict two form factors
(in practice a single one dominates) from non-perturbative
QCD. This is a strong motivation to extend the HQET programme of 
the ALPHA-collaboration\cite{hqet:pap1,hqet:first1,hqet:first2,hqet:first3,
zastat:nf2,hqet:nf2:1} to include matrix elements of all components
of the weak heavy-light currents. And it is a significant step
beyond what has been achieved so far, where only the 
HQET action and the time-component of the axial current were
determined non-perturbatively\cite{hqet:first1,hqet:nf2:1}.

Instead of the previous five we now need \npar 
parameters in order to have the effective theory defined non-perturbatively
including all $1/m$ terms, namely all terms of mass dimension
five in the action and dimension four in the currents. 
Therefore, \npar matching conditions
are needed. It is important to choose them well. Each matching condition
simply consists of a matching observable $\Phi_i$ 
which is evaluated in QCD and in
HQET --- in the latter theory including the terms of order $1/m$
and no more.
Setting $\Phiqcd_i=\Phihqet_i$ determines (in fact defines)
the parameters in HQET. What does it mean to choose the 
matching observables well? Ideally we would like each one of
them to be sensitive
to a single parameter in HQET, in practice  we would like them to receive little 
contributions from terms of order $1/m^2$ in the effective theory. 
If such contributions from $\rmO(1/m^2)$ terms are unnaturally large, they 
affect the determined parameters and then inflict unnaturally large $1/m^2$
terms into the observables that one wants to determine from HQET after 
the matching has been carried out.
One thus better chooses the matching observables in QCD which are strongly 
dominated by the terms of order $m^0$ and $m^{-1}$. Since the ALPHA strategy 
consists of matching in a finite volume with \SF\ boundary conditions,
the size of different terms in the expansion is given in terms of
$z^{-n}=(Lm)^{-n}$ with $L$ the linear extent of the finite volume.

Of course, in the whole process, the most important terms are
those which appear at order $m^0$, the static terms. They are simply
dominating numerically.
It is thus of importance
to make sure that those matching observables which determine the
normalization of 
the static currents are chosen well. Due to the breaking of relativistic 
invariance we need to normalize  the space and time components
of the currents separately. Thus we consider the axial vector current $A_0,A_k$
and the vector one  $V_0,V_k$. Previously, the normalization factor
$\zahqet$ of $A_0$ has been studied in 
detail~\cite{zastat:pap1,zastat:pap1,zastat:pap2,zastat:pap3,zastat:nf2,hqet:first1,hqet:nf2:1}. It
is defined through a \SF\ two-point function \cite{hqet:first1}. Since
in static approximation $A_0$ and $V_k$ are related through the spin symmetry
(see \sect{s:hqet} for a more precise statement), the natural condition
for $\zvkhqet$ follows from a simple spin rotation. However, $\zakhqet$
and $\zvhqet$ do not appear in the \SF\ two-point functions which have been
considered so far. We are thus lead to either consider two-point functions
with more complicated kinematics or three-point functions. 

In fact three-point functions appear naturally, since they are 
also used to determine 
the desired form factor for $B\to\pi l \nu$\cite{
  Bailey:2008wp,Dalgic:2006dt,Liu:2011raa,Bahr:2012qs,Zhou:2012sna,Bouchard:2012tb,Kawanai:2012id
}.
One thus uses a process in the finite volume matching 
which is related to one of the desired infinite volume matrix elements 
and there is even a
potential that higher order in $1/m$ terms cancel between the matching and the
physical matrix element. On the other hand, these functions have not been 
considered before. We therefore evaluate them first in perturbation theory,
including the one-loop parts. We can then verify that they are indeed 
dominated by the first two terms in the $1/m$-expansion. 

The perturbative study is rather straight forward, since one of us has 
developed ``\pastor'', a tool to carry out one-loop computations of \SF\ correlation
functions in a largely automatic manner. Still, the scope of this paper 
is not to consider the full
system of 19 unknowns, but to study the two numerically dominating 
matching conditions for
 $\zvhqet$ and $\zakhqet$.
The \pastor\ software
package was first introduced in \cite{lat11:dirk} and the
publication of a more thorough description along with the source code
is planned for the near future.

\section{The large mass limit of QCD: Heavy Quark Effective Theory
\label{s:hqet}}

\newcommand{\cm}{\ensuremath{\mathcal{M}}}
\newcommand{\cmqcd}{\ensuremath{\mathcal{M}^\mathrm{QCD}}}

We consider QCD with at least three flavors, 
one of them
massive, $m_\beauty=m$, and the others massless, 
in particular $m_\up=m_\down=0$.  A pseudo-scalar state with the flavor 
content $\beauty\bar\down$ is written $|P_{\beauty\bar\down}\,,\;L\rangle$,
with $L$ denoting a single external (kinematical) length scale. Analogously
a light pseudo-scalar state is $|P_{\up\bar\down}\,,\;L\rangle$ and
vector states are labelled with $V$ instead of $P$. We are interested 
in matrix elements
\bes
   \cmqcd(L,\mbar) = 
   \langle X_{\up\bar\down},L | \hat J_\nu^{\up\beauty}(\vecx) |  
           X_{\beauty\bar\down},L\rangle\,,\quad 
   \label{e:matrixel}
\ees
of the QCD heavy-light current operators which correspond
to the classical field 
\bes
  J_\nu^{\up\beauty}(x) = Z_J\, \psibar_\up(x) \Gamma_\nu \psi_\beauty(x)\,. 
\ees
In particular we consider the axial vector current, $J_\nu=A_\nu$,
with $\Gamma_\nu=\gamma_5\gamma_\nu$ and  the vector current, $J_\nu=V_\nu$
with $\Gamma_\nu=\gamma_\nu$. In physical processes, $L$ is an
inverse momentum scale, but we will later use states 
in a finite periodic $L\times L \times L$ volume. For the moment 
the relevant point is that $L$ is the only scale apart from
$m$.  Then there is a perturbative expansion
\bes
\cmqcd(L,\mbar) = ({\cmqcd})^{(0)}(z) + \gbar^2(L) ({\cmqcd})^{(1)}(z) 
             + \rmO(\gbar^4(L)) \,,\quad z=L\mbar \,.
\ees
We will specify the renormalization scheme for $\gbar,\mbar$ 
when it becomes relevant. 
The renormalization factors $Z_J$ of the flavor currents are to be chosen 
such that the currents satisfy the chiral 
Ward identities\cite{curralgebra:MaMA,impr:pap4}. In the 
large mass limit, $\mbar\to\infty$, $L$ fixed, the matrix elements $\cmqcd$
are logarithmically divergent \cite{Shifman:1987sm,Politzer:1988wp},
\bes
   ({\cmqcd})^{(1)}(z) \simas{z\to\infty}\; \Hone - \gamma_0 \log(z) \Hzero \,,
   \quad
   \gamma_0=-1/(4\pi^2)\,,\quad z=L\mbar\,.  \label{e:asym}
\ees  
This limit of QCD is described by an effective field theory, 
HQET.
Up to corrections of order $1/z$, it is 
the static effective theory~\cite{stat:eichhill1} where 
the b-field is replaced by a two-component static field, 
\bes
   \psi_\beauty(x) \to \heavy(x)=\frac12(1+\gamma_0)\heavy(x)\,,
\ees
with Lagrangian\footnote{We are
in the frame where  $|X_{\beauty\bar\down}\rangle$ has spatial momentum zero
and HQET at zero velocity applies.},
\bes
  \Lstat(x) = \heavyb(x) (\dmstat + D_0) \heavy(x) \,.
\ees
The mass counter term $\dmstat$ does not play a role in the following.
The static flavor currents are form-identical with the QCD ones, for example  
$\Vstat(x) = \psibar_\up(x) \gamma_0 \heavy(x)$, 
$\Akstat(x) = \psibar_\up(x) \gamma_5\gamma_k \heavy(x)$. 
Chiral Ward identities fix the relative normalization of the static
vector and axial vector currents but not the overall normalization.
Furthermore space and time-components
are to be treated separately and the currents have an anomalous 
dimension in the effective theory.  Choosing the lattice regularization
we can in a first step define finite currents by renormalizing them
in the lattice minimal subtraction scheme. The renormalized currents 
are then
\bes
  (J_\mathrm{lat}^\mathrm{stat})_\nu(x;\mu) = 
  Z_\mathrm{lat}(\mu a, g_0)\,J^\mathrm{stat}_\nu(x) =
  Z_\mathrm{lat}(\mu a, g_0)\,\psibar_\up(x) \Gamma_\nu \heavy(x)\,, 
\ees
with a renormalization constant
\bes
  Z_\mathrm{lat}(\mu a, g_0) = 1 - \gamma_0 \log(a\mu) g_0^2 + \rmO(g_0^4)\,, 
\ees
which is common to all currents (see \cite{LH:rainer} for a pedagogical 
introduction). Their matrix elements 
\bes
   \cm^\mathrm{stat}_{J_\nu}(L,\mu) = Z_\mathrm{lat}(\mu a, g_0)
   \langle X_{\up\bar\down} | \hat J^\mathrm{stat}_\nu(\vecx) |  
           X_{\beauty\bar\down}\rangle_\mathrm{stat}\,,
\ees
are then finite. When we set $\mu=\mbar$, they are equal to the 
corresponding QCD matrix elements 
up to higher order terms in $1/m$,
\bes
  \cmqcd_{J_\nu}(L, \mbar) = C_{J_\nu}^\mathrm{match}(\gbar^2(\mbar))\, 
   \cm^\mathrm{stat}_{J_\nu}(L,\mbar) + \rmO(1/\mbar) \,, 
  \label{e:match}
\ees
and up to the finite renormalization factor 
\bes
  C_{J_\nu}^\mathrm{match}(g^2) =  1+ B_{J_\nu}g^2 + \rmO(g^4) \,.
\ees  
The one-loop coefficients are
\bes
   B_{A_0} &=& -0.137(1)\,, \label{e:Bastat}
   \\
   B_{V_0} - B_{A_0} &=& 0.0521(1) = B_{V_k} - B_{A_k}\,,\label{e:Zva}
   \\
   B_{A_k} - B_{V_0} &=& -0.016900\,. \label{e:diffB}
\ees
Here \eq{e:Bastat}, due to \cite{BorrPitt,zastat:pap2}, and
\eq{e:Zva}, due to \cite{zvstat:filippo}, depend on the 
lattice regularisation. They are given 
for the Eichten-Hill lattice action for the 
static quark, the $\rmO(a)$-improved Wilson action for the light quarks
and the plaquette gauge action. We note that \eq{e:Zva} follows
from requiring a chiral Ward identity. On the other hand the bare currents 
$V_0$ and $A_k$ are related by the spin symmetry of the static effective 
theory which is exact in lattice regularization. The difference,
\eq{e:diffB}, is therefore known very precisely from continuum perturbation
theory~\cite{BroadhGrozin2}. Of course the renormalization of the
fields and therefore in particular $B_{J_\nu}$ are independent of the states
in \eq{e:matrixel}.

\section{Matching conditions}

\subsection{Definitions of correlation functions}

As discussed in the introduction, in the ALPHA strategy 
we use finite volume matrix elements
to define the matching of HQET and QCD. These matrix elements
are constructed in the \SF, where they are exactly related 
to ratios of correlation functions, see \cite{hqet:test1} for more details.  
Here we define those correlation functions and ratios which
are suitable for the matching of $V_0$ and $A_k$. 

We choose the \SF\ with vanishing background field, denote the time-extent
by $T$ and the space-extent by $L$.
As a shorthand we introduce (non-local) boundary fields
\bes
   \Obound{ij}{\Gamma} &=& {a^6\over L^3}
         \sum_{\vecx,\vecy} 
         \zetabar_i(\vecx) \Gamma \zeta_j(\vecy)\,,
   \quad
   \Oboundp{ij}{\Gamma} = {a^6\over L^3}
         \sum_{\vecx,\vecy} 
         \zetabarprime_i(\vecx) \Gamma \zeta'_j(\vecy)\,,   
\ees
where the first one creates a meson with flavor content $i\bar j$ 
at time zero and the second annihilates a meson with flavor content
$j\bar i$ at final time $T$. The boundary quark fields 
$\zeta_i,\zetabar_i$ are defined in \cite{impr:pap1}.
For simplicity and because more sophisticated choices seem
unnecessary, we take each flavor to have the same 
periodicity phase $\theta$ in the boundary conditions
$\psi(x=L\hat k) = \rme^{i\theta} \psi(x)\,,\; 
 \psibar(x=L\hat k) = \rme^{-i\theta} \psibar(x)$~.

With these preliminaries we define boundary-to-boundary correlation functions
(remember $z=\mbar L$)
\bes
    \fone^{\beauty\down}(\theta,z) &=& -\frac12
         \langle \Oboundp{\down\beauty}{\gamma_5}
         \Obound{\beauty\down}{\gamma_5} \rangle \,,
    \\
    \fone^{\up\down}(\theta) &=& -\frac12
         \langle \Oboundp{\down\up}{\gamma_5}
         \Obound{\up\down}{\gamma_5} \rangle \,,
    \\
    \kone^{\up\down}(\theta) &=& -\frac12 
         \langle \Oboundp{\down\up}{\gamma_k} 
         \Obound{\up\down}{\gamma_k} \rangle \,,
\ees
and three-point  correlation functions with the desired 
currents
\bes
    \fv(x_0;\theta,z) &=& -\frac{L^3}2
         \langle \Oboundp{\down\up}{\gamma_5}V_0^{\rm ub}(x)
         \Obound{\beauty\down}{\gamma_5} \rangle\,,
    \\
    \ja(x_0;\theta,z) &=& -\frac{L^3}2
         \langle  \Oboundp{\down\up}{\gamma_1} A_1^{\rm ub}(x)
         \Obound{\beauty\down}{\gamma_5} \rangle\,.
\ees

\subsection{Possible matching observables for $V_0,A_k$}

The defined correlation functions are easily combined to form the desired
finite volume matrix elements,
\bes
   L^3 \cmqcd_{\rm V_0}(L,\mbar) &=& 
   - \zv\,{{\fv(T/2;\theta,z) \over 
   [\fone^{\up\down}(\theta) \fone^{\beauty\down}(\theta,z)]^{1/2}}}\,,
   \label{e:phiv0}
   \\
   L^3 \cmqcd_{\rm A_{k}}(L,\mbar) &=& 
   - \za\,{{\ja(T/2;\theta,z) \over 
   [\kone^{\up\down}(\theta) \fone^{\beauty\down}(\theta,z)]^{1/2}}}\,, 
   \label{e:phiak}
\ees
where we set $ T=L$. As explained in \cite{hqet:test1} these ratios 
are equal to the matrix elements \eq{e:matrixel} with the 
finite volume states such as 
$|P_{\beauty\bar\down}\,,\;L\rangle$, all normalized to unity. 
We here neglect $\Oa$-improvement, but this is used in the perturbative
computations in \sect{s:pt}.

We now have good candidates for matching conditions which we write in
the form
\be
  \Phiqcd_{J_\nu}(L, \mbar) = \Phi^\mathrm{stat}_{J_\nu}(L, \mbar) 
  + \log \left\{
  C_{J_\nu}^\mathrm{match}\left(\gbar^2(\mbar)\right)\right\} + \rmO(1/\mbar)\,,
\label{e:matchphi}
\ee
with $\Phiqcd_{J_\nu} \equiv \log\left(L^3\cmqcd_{J_\nu}\right)$. In this
way the $\log(C_{J_\nu}^\mathrm{match})$-term appears additively, which is advantageous
once the $1/m$-terms are included \cite{hqet:first1}.

\subsection{Checking their quality}

Expanding \eq{e:matchphi} in the coupling we have 
\bes
  (\Phiqcd_{J_\nu})^{(0)}(z) &=&  
    (\Phi^\mathrm{stat}_{J_\nu})^{(0)} + \rmO(1/z) \,, 
  \\
  (\Phiqcd_{J_\nu})^{(1)}(z) &=& 
   (\Phi^\mathrm{stat}_{J_\nu})^{(1)} +  B_{J_\nu} - \gamma_0 \log(a \mbar)  
           + \rmO(1/z)\,.
\ees
The one-loop part can be rewritten as in \eq{e:asym}, namely
\bes
  \Gone_{J_\nu}(z) &\equiv& (\Phiqcd_{J_\nu})^{(1)}(z) 
      + \gamma_0 \log(z) = \Hone_{J_\nu} + \rmO(1/z)
  \label{e:oneloopsubtr}
\ees
with 
\bes
  \Hzero_{J_\nu}&=&(\Phi^\mathrm{stat}_{J_\nu})^{(0)} \,,
  \\
  \Hone_{J_\nu}&=& (\Phi^\mathrm{stat}_{J_\nu})^{(1)} +  B_{J_\nu} - \gamma_0  \log(a/L)
          \,,
\ees
where we subtract the logarithmic singularity
in $z$ from $(\Phiqcd_{J_\nu})^{(1)}(z)$ such that $\Hone_{J_\nu}$ represents the one-loop
coefficient of the matched static matrix element at renormalization scale $1/L$. 
In this form the size of $1/m$ terms is directly 
visible as deviations of the left hand side of \eq{e:oneloopsubtr} from
$\Hone_{J_\nu}$. We want to investigate these deviations in the following
in order to ensure that \eq{e:phiv0} and \eq{e:phiak}  are good observables 
for the matching.



\newcommand{\Zm}{Z_\mathrm{m}}
\newcommand{\ord}[1]{^{(#1)}}

\newcommand{\foud}{\ensuremath{F_{\mathrm 1}^{\mathrm{ud}}}}
\newcommand{\koud}{\ensuremath{K_{\mathrm 1}^{\mathrm{ud}}}}
\renewcommand{\fv}{\ensuremath{F_{\mathrm{V_0}}}}
\renewcommand{\ja}{\ensuremath{J_{\mathrm{A_1}}^{\mathrm 1}}}
\renewcommand{\ga}{\ensuremath{G_{\mathrm{A_k}}}}
\newcommand{\gv}{\ensuremath{G_{\mathrm{V_0}}}}
\newcommand{\hv}{\ensuremath{H_{\mathrm{V_0}}}}

\renewcommand{\mbar}{\ensuremath{\overline{m}}}
\newcommand{\gzero}{\ensuremath{{g}_0}}
\newcommand{\gzerosq}{\ensuremath{{g}_0^2}}
\renewcommand{\zmlat}{\ensuremath{Z_{\mathrm{m},\lat}}}

\section{One-loop computation \label{s:pt}}

All the required quantities (\fv, \ja, \foud, \koud, and their static
counterparts) were calculated at the one-loop level using the \pastor\
software package for automated lattice perturbation theory
calculations~\cite{lat11:dirk}. As input, \pastor\ accepts a rather general class of
lattice actions and observables defined in the Schr\"odinger
functional. It will then automatically generate computer programs for
the evaluation of all contributions of the observables under
investigation up to one-loop order including improvement- and
counter-terms. We did implement full $\Oa$-improvement, including
the terms proportional to $a\mq$ not written in \eq{e:phiv0} and
\eq{e:phiak}.

\begin{figure}[hb]
  \centering
  \includegraphics{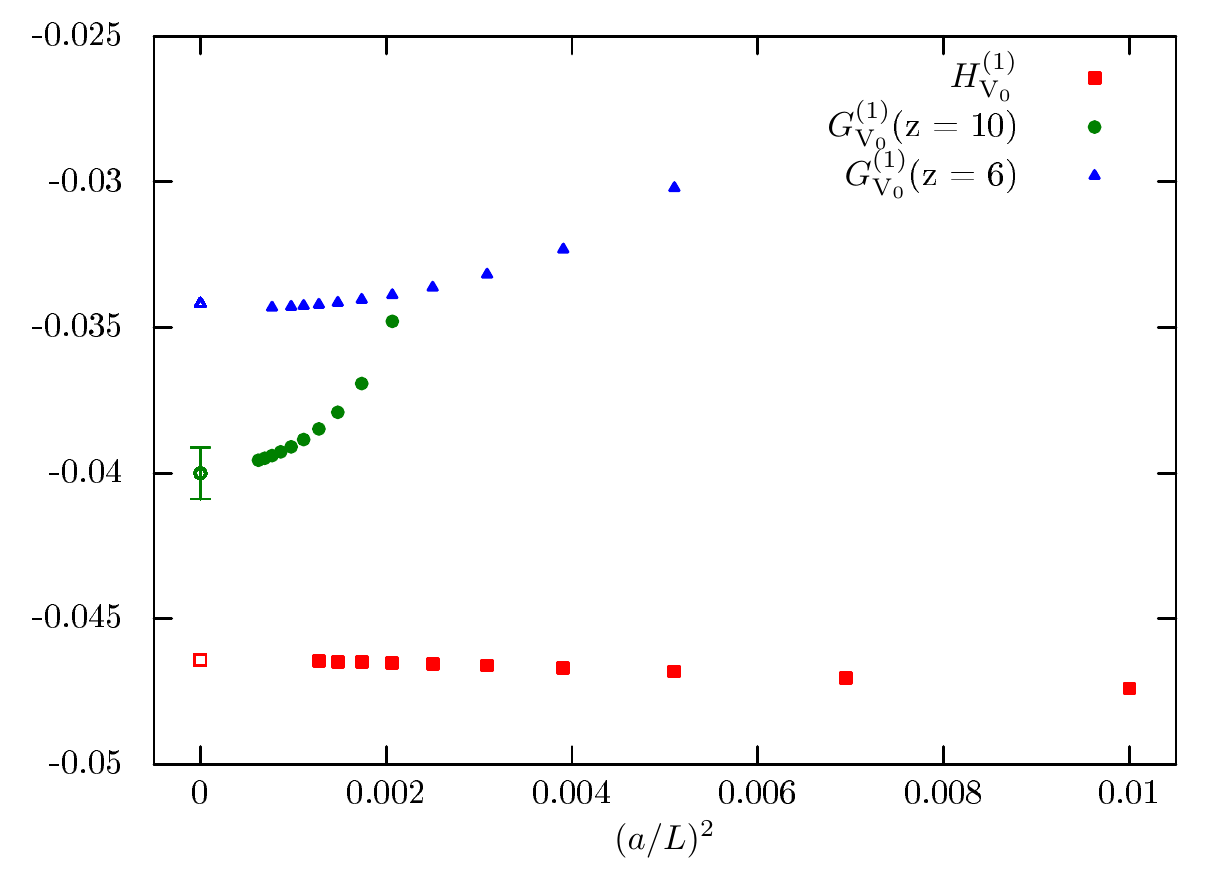}
  \caption{Continuum extrapolation of \hv, \gv
    at one-loop level, $\theta = 0.5$. The round-off errors on the
    data points at finite $L/a$ and the uncertainty of the continuum
    extrapolation for the static point are much smaller than the
    symbol size. 
    }
  \label{fig:cont_etr}
\end{figure}

For the quantities in QCD, we choose lattice resolutions of $L/a$ up
to 40, while for the HQET counterparts lower resolutions up to $L/a =
30$ are sufficient to obtain reliable continuum extrapolations,
c.f. \fig{fig:cont_etr}. To determine the continuum limits, we
employ the method described in \cite{pert:2loop_fin} using the
implementation provided by \pastor. We choose $\theta \in \{0, 0.5, 1.0\}$
and $z \in\{ 4, 6, 8, 10\}$. 

We employ the
mass-independent lattice minimal subtraction scheme \cite{impr:pap1}
in which the $\Oa$ improved renormalized mass at scale 
$\mu=1/L$ is given by 
\begin{equation}
  \label{eq:3}
  \mbar(L) = \zmlat(\gzerosq, a/L) \, \mq \left[ 1 + a \,\bm(\gzerosq) \,
    \mq\right], \quad \mq = m_0 - \mc\,
\end{equation}
in terms of the bare mass of the lattice theory. At one-loop order
we have \cite{impr:pap5,pert:gabrielli} 
\begin{align}
  \bm(\gzerosq) \;&= - 0.5 - 0.07217(2)\,C_F\, \gzerosq + O(\gzero^4),\\
  \zmlat(\gzerosq, a/L) \;&= 1 - \frac  1 {2\,\pi^2} \log (a/L)
     \gzerosq + O(\gzero^4).
\end{align}
All calculations in \pastor\ are performed with $z = \mbar(L) L$ as
input. It inverts \eq{eq:3}  to obtain
$m_0 = m_0 \ord 0 + \gzerosq m_0 \ord 1 + O(\gzero^4)$ and calculates the series
\begin{multline}
  \label{eq:4}
  {\cal O} \left(m_0 \ord 0 + \gzerosq m_0 \ord 1\right) = {\cal
    O}\ord 0\left(m_0 \ord 0\right) \\+ \gzerosq \left[ {\cal O}\ord
    1\left(m_0 \ord 0\right) + m_0 \ord 1 \partial_{m_0} {\cal O}\ord
    0\left(m_0 \ord 0\right) \right] + O\left(\gzero^4\right)
\end{multline}
for a given observable ${\cal O}(m_0)$.
For the evaluation of the diagrams of a Schr\"odinger functional
observable, it is beneficial to work in a time-momentum
representation. Due to the periodic spatial boundary conditions one
does not have to perform a momentum-integration but a sum over a
finite set of allowed lattice momenta of size $(L/a)^3$. The round-off
errors introduced by the numerical evaluation of this sum are
estimated from the difference of \texttt{long double} precision and
\texttt{double} precision results for representative
parameters. Apart from this test we use \texttt{double} precision
since it is roughly a factor three faster.
The execution time to evaluate the numerically most
challenging loop diagram at $L/a = 40$ was about
50 hours on a single core CPU (Nehalem).

\section{Results}
\begin{figure}[htb!]
  \centering
\includegraphics{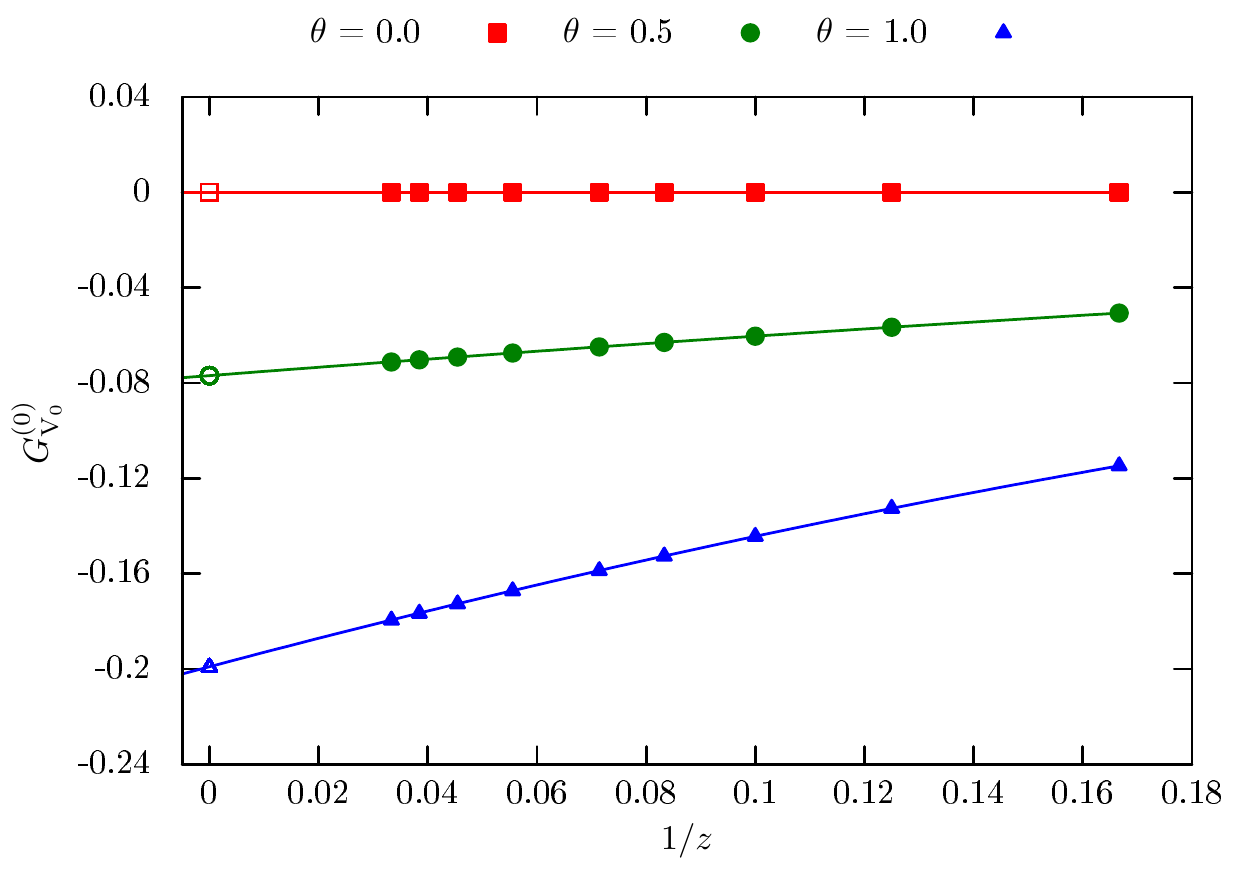}
  \caption{$\Gzero_{V_0} \equiv \left(\Phiqcd_{V_0}\right)^{(0)}(z)$ 
    in the continuum limit. Errors are much smaller
    than the symbol size.}
  \label{f:v0}
\end{figure}
\begin{figure}[htb!]
\begin{center}
\includegraphics{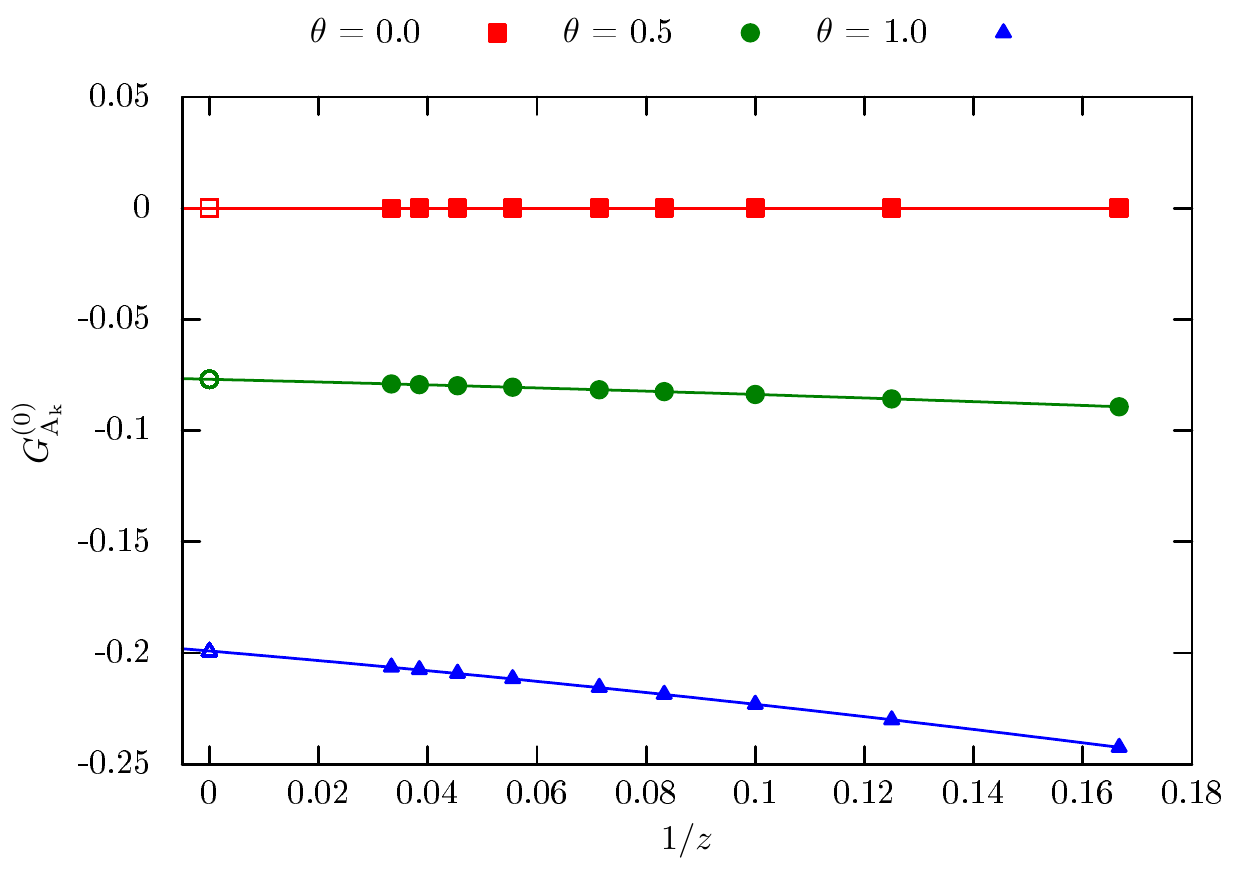}
\end{center}
\caption[]{\label{f:a1}
    $\Gzero_{A_1} \equiv \left(\Phiqcd_{A_1}\right)^{(0)}(z)$ 
    in the continuum limit. Errors are much smaller
    than the symbol size.
}
\end{figure}

\subsection{Tree-level}
We start the discussion of our results with the tree-level functions
$\Gzero_{\rm V_0}(z) \equiv (\Phiqcd_{\rm V_0})^{(0)}(z)$ and 
$\Gzero_{\rm A_1}(z) \equiv (\Phiqcd_{\rm A_k})^{(0)}(z)$. 
Together with the static values $\Hzero_{J_\nu}$ they are displayed in 
\fig{f:v0} and \fig{f:a1} for three different values
of $\theta$.  Curves are fits of the form
$\Hzero(1 + h_1/z +h_2/z^2)$, fitted to the data with weights
$w(z)=1/z^3$. The fits are thus dominated by the results at large
$z$. The coefficients $h_i$, listed for the 
different cases in \tab{t:treelevelzexp}, are small. For all considered values
of $\theta$ the $1/m$-expansion is 
well behaved and we can also be confident that the fitted coefficients
are close to the true Taylor coefficients. Obviously, from the point of view 
of tree-level, one would prefer $\theta=0$ where 
$\Gzero_{J_\nu}(z) = \Hzero_{J_\nu}$ holds exactly. 

\begin{table}[ht!]
  \centering
  \begin{tabular}{c | c c | c c | c c}
    $\theta$ & \multicolumn{2}{c|}{0.0} & \multicolumn{2}{c|}{0.5} &
    \multicolumn{2}{c}{1.0}\\\hline\hline
    & $h_1$ & $h_2$ & $h_1$ & $h_2$ & $h_1$ & $h_2$\\\hline
$G_{\mathrm{A_k}}^{(0)}$& 0.00000 & 0.00000& 0.77621 & 1.11933& 1.05083 & 1.53061\\[1ex]
$G_{\mathrm{V_0}}^{(0)}$& 0.00000 & 0.00000& -2.30791 & 1.57043& -3.06017 & 3.11951\\[1ex]\hline\hline
    & $h_1$ & $f_1$ & $h_1$ & $f_1$ & $h_1$ & $f_1$\\\hline
\multirow{2}{*}{$G_{\mathrm{A_k}}^{(1)}$}& 0.05245 & -0.00132& 0.12513 & 0.00139& 0.21893 & 0.01547\\
& 0.03099 & 0.01054& 0.10547 & 0.01225& 0.19391 & 0.02929\\\hline
\multirow{2}{*}{$G_{\mathrm{V_0}}^{(1)}$}& 0.15093 & -0.00923& 0.08803 & -0.00811& 0.04548 & -0.01340\\
& 0.12100 & 0.00692& 0.07042 & 0.00139& 0.05268 & -0.01728\\\hline
  \end{tabular}
  \caption{Fit coefficients for \ga and
    \gv. The upper row of fit coefficients for the one-loop results comes
    from the fits omitting the data at $z = 4$.}
  \label{t:treelevelzexp}
\end{table}

\subsection{One-loop}
We get more information at one-loop order. In order to have all 
finite pieces defined, we need to specify the renormalization scheme
for the quark mass. As stated in \sect{s:pt}, 
we take $\mbar$ to be the renormalized mass
in the lattice minimal subtraction scheme at scale $\mu=1/L$.
The continuum limit is taken as described in the previous section.

\begin{figure}[htb!]
\begin{center}
\includegraphics{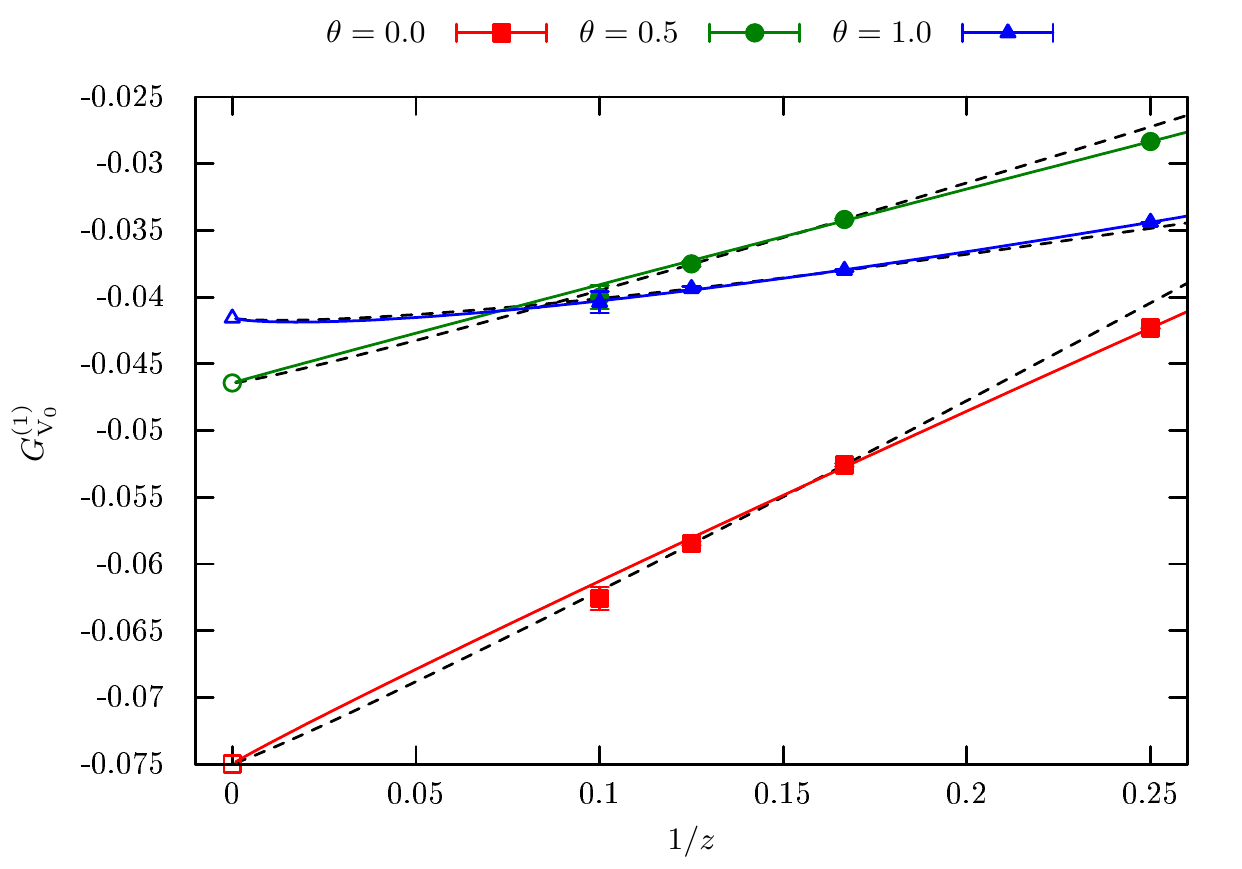}
\end{center}
\caption[]{\label{f:v01lp}
$\Gone_{\rm V_0}(z)$ in the continuum limit,
compared to the static result.  
}
\end{figure}

\begin{figure}[htb!]
\begin{center}
\includegraphics{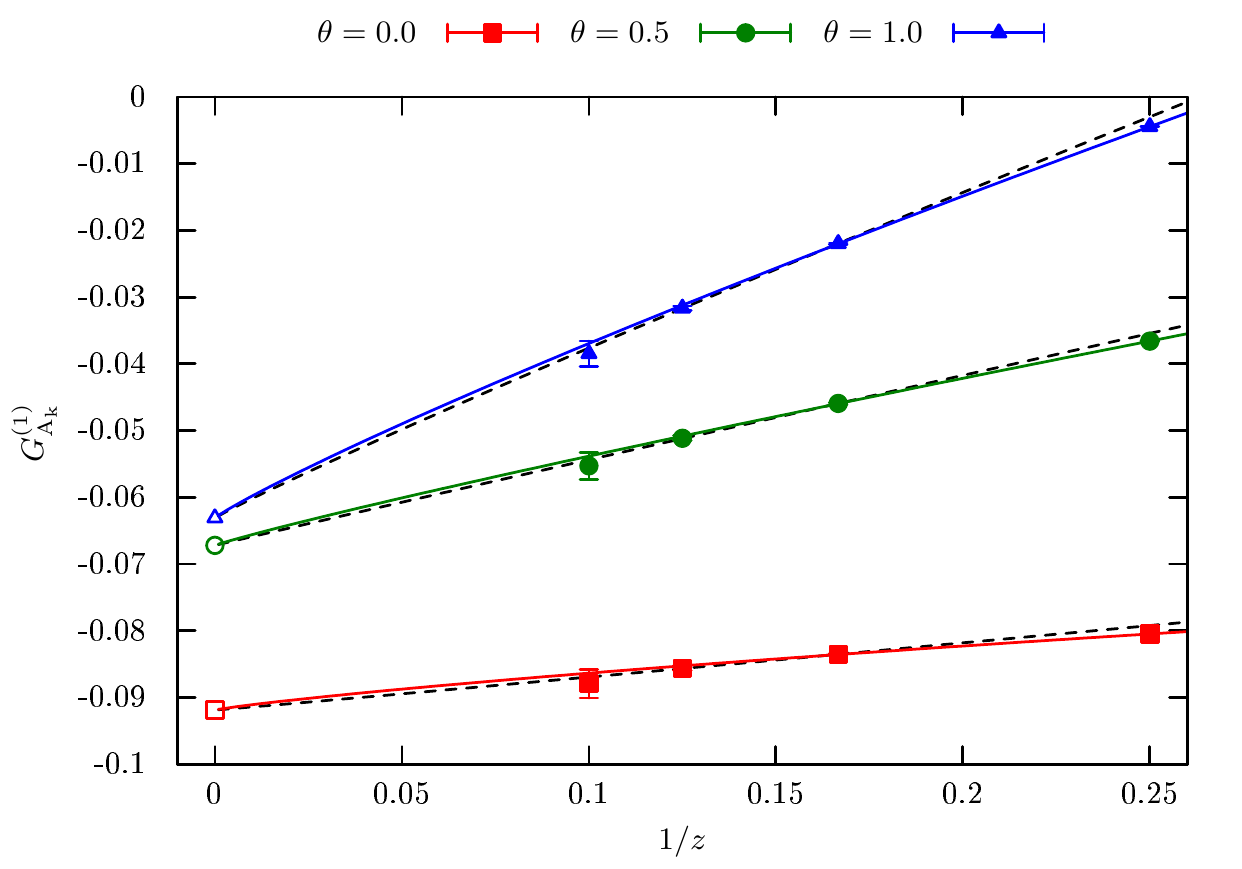}
\end{center}
\caption[]{\label{f:a11lp}
$\Gone_{\rm A_1}(z)$ in the continuum limit,
compared to the static result.  
}
\end{figure}

The combination $G_{J_\nu} \ord 1 (z)$, \eq{e:oneloopsubtr}, is shown in \fig{f:v01lp}
and \fig{f:a11lp}. We perform a fit to the one loop data employing a
function of the form 
\begin{equation}
  \label{e:oneloopfit}
  G_{J_\nu} \ord 1 (z) = H_{J_\nu}\ord 1 + h_1 /z + f_1 \log(z)/z,
\end{equation}
choosing in this case constant weights, as only few data point are
available anyway. It is compared to a fit of the same form, omitting
the data at $z = 4$. The fit parameters for the one-loop quantities in
\tab{t:treelevelzexp} are not expected to be accurate estimates for
the corresponding asymptotic expansion. The accuracy of the fits and
the smallness of the coefficient $f_1$, however, may be taken as an
indication that higher order terms in the $1/z$-expansion are not
very important for the considered range in $z$.

The size of $\Hone$ is relevant for us only as a consistency check:
for all cases it is a little smaller than the expected magnitude $1/(4\pi)$
for a perturbatively accessible quantity. The interesting question is the 
magnitude of $1/z$-terms as well
as curvature when $G_{J_\nu}$  are considered a function of $1/z$.

We observe that also at one-loop order 
the $1/z$-terms in $\Phi_{J_\nu}$ remain small, but $\theta=0$ 
is not preferred any more. A choice $\theta=0.5$ appears a good compromise
between tree-level and one-loop. Take for illustration $\gbar^2=4$ and 
$z \geq 10$ 
as it is typical in the non-perturbative application \cite{hqet:nf2:1}.
Then we roughly have a few per-mille $1/z$ correction at tree-level
and an $\approx 3\%$ correction at one-loop. This is very acceptable.
We then have all rights to expect that the $1/z^2$ corrections, which are omitted 
when HQET is treated non-perturbatively \cite{hqet:first3,lat11:patrick}, 
are negligible and indeed the curvatures seen in \fig{f:v01lp} and \fig{f:a11lp}
are small.

\clearpage
\section{Conclusions}
The proposed three-point functions appear very useful. 
They are seen to be strongly dominated by the lowest terms in the $1/z$ 
expansion. As a consequence, the three-point functions may well be applied 
to fix the remaining two unknowns,  $\zvhqet$ and $\zakhqet$, in the
static approximation  non-perturbatively. We would recommend $\theta=0.5$,
but the one-loop study does not suggest this choice to be much superior to
$\theta=0$ or $\theta=1$. 
At order $1/m$ the full system determining the 19 parameters has to be
considered. Three of these parameters come from the HQET action
\cite{stat:eichhill2}, two from the temporal components of the vector
and axial vector current respectively and the spatial components
of the currents require the inclusion of further six parameters each
\cite{Falk:1990de,Falk:1992fm,Neubert:1992tg}. A study of this system
in perturbation theory is presently being carried out by the ALPHA
collaboration.

We can also confirm that the new package \pastor\ is very useful in
studying such problems in perturbation theory. This goes beyond 
issues related to the regularization such as renormalization factors
or improvement coefficients. In fact, all results 
presented here refer to the $z$-dependence in continuum perturbation theory, 
since we were able to reliably take the continuum limit
$a/L\to0$. We have presented the results in the 
lattice minimal subtraction scheme for the quark mass. They can
trivially be connected to the $\msbar$ scheme by using 
\cite{pert:gabrielli}
$\mbar(L)=(1+0.122282\,\gbar^2) \times \mbar_\msbar(1/L) +
\rmO(\gbar^4)$.\\[1em]
{\noindent \bf Acknowledgements.} We want to thank Piotr Korcyl,
Michele della Morte and Hubert Simma for helpful discussions, Jochen
Heitger for a critical reading of our draft and the computing center
at DESY Zeuthen for support and CPU time on the PC farm. This work has
been partly funded by the Research Executive Agency (REA) of the
European Union under Grant Agreement number PITN-GA-2009-238353 (ITN
STRONGnet) and by the SFB/TR 9 of the Deutsche Forschungsgemeinschaft.

\eject

%
  \bibliography{latticen,HQET,my}        
  \bibliographystyle{JHEP}   

\end{document}